\documentclass[aps,prb,showpacs,twocolumn]{revtex4}
\usepackage{graphicx}
\usepackage{amssymb}
\usepackage{rotating}

\newcommand{\?}{\hspace{-1mm}}
\newcommand{\be}{\begin{equation}}  \newcommand{\ee}{\end{equation}}
\newcommand{\bea}{\begin{eqnarray}}  \newcommand{\eea}{\end{eqnarray}}
\newcommand{\vk}{\mathbf k} \newcommand{\vK}{\mathbf K}
\newcommand{\vx}{\mathbf x} \newcommand{\vvr}{\mathbf r}
\newcommand{\blx}{\rule{1.2mm}{1.2mm}}

\begin{document}
\title{Energy gap in graphene nanoribbons with structured external electric potentials}
\author{W. Apel}
\author{G. Pal}
\author{L. Schweitzer}
\affiliation{Physikalisch-Technische Bundesanstalt (PTB),
Bundesallee 100, 38116 Braunschweig, Germany}
\begin{abstract}
The electronic properties of graphene zig-zag nanoribbons with
electrostatic potentials along the edges are investigated. Using the
Dirac-fermion approach, we calculate the energy spectrum of an
infinitely long nanoribbon of finite width $w$, terminated by
Dirichlet boundary conditions in the transverse direction. We show
that a structured external potential that acts within the edge
regions of the ribbon, can induce a spectral gap and
thus switches the nanoribbon from metallic to insulating behavior. 
The basic mechanism of this effect is the selective influence of the
external potentials on the spinorial wavefunctions that are
topological in nature and localized along the boundary of the
graphene nanoribbon. Within this single particle description, the
maximal obtainable energy gap is $E_{\rm max}\propto \pi\hbar v_{\rm
  F}/w$, i.e., $\approx 0.12$\,eV for $w=$15\,nm. The stability of the
spectral gap against edge disorder and the effect of disorder  on the
two-terminal conductance is studied numerically within a tight-binding
lattice model. We find that the energy gap persists as long as the
applied external effective potential is larger than $\simeq 0.55\times
W$, where $W$ is a measure of the disorder strength. We argue that
there is a transport gap due to localization effects even in the
absence of a spectral gap. 

\end{abstract}
\date{\today} 
\maketitle

\section{Introduction}
The continuing rise of graphene as a new and exceptionally promising
material that outperforms conventional metals and semiconductors has
initiated an ongoing quest for new physical effects. This has also
generated a multitude of exciting proposals for various technical
applications, which have recently been summarized in several 
reviews.\cite{CGPNG08,Bee08,Cas10} However, due to single
layer graphene's gap-less energy structure, applications 
where considerable on-off current ratios are indispensable are
limited at present. Proposals for the creation of a lattice anisotropy 
that would lift the sublattice symmetry,\cite{GKBKB07} or for the
application of strain fields \cite{GLZ08,PCP09} that also could open an
energy gap have yet to be realized. At present, bilayer graphene or
certain graphene arm-chair nanoribbons have to be utilized instead, if 
an energy gap is needed. In the latter case, narrow ribbons of
special widths have to be fabricated so that, due to quantum
confinement, an energy gap or at least a transport gap in disordered
ribbons is formed. For ribbon widths below 30\,nm, the spectral gap 
is larger than $kT$ at room temperature.\cite{HOZK07,LPCA08} Recently,
the effect of a transversal electric field on arm-chair ribbons has
also been studied theoretically.\cite{Nov07}  

Within simple non-interacting particle descriptions, graphene zig-zag
ribbons are metallic and the opening of a spectral gap is impeded by
electronic edge states \cite{FWNK96,NFDD96,WTYS09} appearing in an
energy range where for broad two-dimensional graphene sheets valence
and conduction bands touch. These edge states are sensitive to an
Aharonov-Bohm flux and robust against edge
reconstructions.\cite{SMS06,SSS08} For interacting electrons,
it was shown \cite{SCL06a} that zig-zag ribbons always have a gap due
to edge magnetization and that a homogeneous external electric field
applied across the ribbon causes a half-metallic state.\cite{SCL06} 
By employing the \textit{ab initio}
pseudopotential density functional method,\cite{Soler02} the authors
of Ref.~\onlinecite{SCL06} studied the spin-resolved electronic 
structure of zig-zag graphene nanoribbons and the possibility of
spin-polarized currents. The influence of electron transfer between
the two edges on the half-metallicity of the nanoribbon subjected
to the electric field was also investigated.\cite{KLYH07} Recently,
a gapped magnetic ground state has been suggested to be due to an
antiferromagnetic interedge superexchange.\cite{JPM09}  

These advanced theories are extremely interesting for clean graphene
zig-zag nanoribbons. However, there is still no conclusive direct
experimental observation proving the existence of an one-dimensional
magnetic state. The latter may well be spoiled in reality by edge
disorder or adsorbent atoms.\cite{KOQF10} Therefore, we try  to
clarify in this paper whether one can obtain a spectral gap already
within a single particle description. 
In order to substantiate the relevance of such a basic model
for real graphene zig-zag nanoribbons, we investigate the influence of
edge disorder on both the spectral properties and the two-terminal
electronic transport. 
In our work, we first investigate a very simple spinless continuum
model that can be treated analytically and then we employ a
tight-binding lattice model including edge disorder effects, which we
solve numerically. 

Based on the Dirac equation for two-dimensional electrons with Dirichlet  
boundary conditions imposed in the transverse direction, we study in
section~\ref{sectlytic} the 
influence of an effective potential acting within narrow strips along
the edge regions of the graphene zig-zag nanoribbon. This set-up, as
sketched in Fig.~\ref{pots}, can be studied experimentally in a
three-dimensional device by applying voltages between the back-gate of
a graphene zig-zag ribbon and two top gates, one at the right and the
other one at the left edge, respectively. 
The results are given in section~\ref{sectresult}. 
For a perfectly antisymmetric electric potential (left side $V$ and 
right side $-V$, see Fig.~\ref{pots}), we find a symmetric spectral
gap at the Dirac point, which increases linearly with the applied
voltage $V$. Increasing the potential further, the gap reaches a
maximum value of $\sim\pi/w$, where $w$ is the width of the ribbon,
and finally closes again for even larger $V$. The splitting of the
edge state energies, however, still continues to rise with $V$.   

In order to check how these results are influenced by disorder, we
study numerically a tight-binding lattice model in
section~\ref{sectdis}, calculate the two-terminal conductance, and 
investigate the influence of edge disorder on both the spectral and
the transport gap. The former survives for $V$ being larger than
$0.55\times W$, where $W^2/3$ is the 2nd moment of the distribution
$P(\epsilon)=1/(2W)\Theta(W-|\epsilon|)$ that defines the disorder
potentials assumed along the ribbon edges. A transport gap, however,
is still observable even for very strong disorder when the spectral
gap is absent.       

\section{Model and solution\label{sectlytic}}
\begin{figure}
\begin{center}
\includegraphics[width=8.0cm]{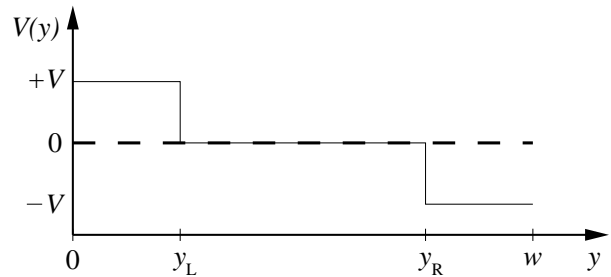}
\caption{Cross section of a graphene zig-zag nanoribbon of width $w$
  (dashed line) showing the effective potential $V(y)$  created by
  external voltages applied between the back-gate and two top-gates. 
  The top-gates extend in the $x$ direction along the left and the
  right edge having widths $y_L$ and $w-y_R$, respectively.} 
\label{pots}
\end{center}
\end{figure}

We study a zig-zag nanoribbon of graphene, infinitely extended in
$x$-direction and with finite width $w$ in $y$-direction (see
Fig.~\ref{pots}) using the Dirac-type equation approximation. This
Dirac-fermion approach usually describes correctly the low-lying
states around the neutrality point in 
graphene.\cite{ANS98,KNG06,GS05,And07,Bee08}   
There are two inequivalent (Dirac) points at $\vk = \pm \vK$ (valleys)
in the band structure where valence band and conduction band touch.
The wave function for wave vectors near $\vK$ is written as
$(\xi(a,\vx), \xi(b,\vx))$, where $a$ and $b$ denote the two sublattices 
of the graphene structure. Correspondingly, $(\eta(a,\vx),
\eta(b,\vx))$ is the wave function in the other valley. 
Since the Hamiltonian does not mix the valleys near $\vK$ and $-\vK$, the 
Dirac equation separates and reads for the first valley 
(in the other valley, we have a corresponding equation with 
$i\partial_x \to -i\partial_x$)
\begin{equation}
\left(\begin{array}{cc}
E-V(y) &  \hbar v_{\mathrm F}(i\partial_x -\partial_y) \\ 
 \hbar v_{\mathrm F}(i\partial_x + \partial_y) & E-V(y)
\end{array}\right)
\left(\begin{array}{cc}\xi(a,\vx) \\ \xi(b,\vx)\end{array}\right) =0.
\end{equation}
Here, $v_{\mathrm F}\simeq 10^{6}$\,m/s is the Fermi velocity, $E$
the energy, and $V$ the electrostatic potential depending only on the 
$y$-coordinate. In what follows, we set $\hbar v_{\mathrm F}=1$ but
recover the units when showing our results in the figures. 
In the corresponding lattice model, 
the boundaries at $y=0$ and $y=w$ are considered to be of
zig-zag type. Then, in a description in terms of the Dirac model, we
have periodic boundary conditions in the $x$-direction and Dirichlet
boundary conditions in the $y$-direction such that $\xi(a,x,y=0)=0$
and $\xi(b,x, y=w)=0$.\cite{BF06a} 

In order to have a simple model that can be treated analytically,
we consider a piecewise constant electrostatic potential $V(y)=V$ for
$0\le y \le y_L$, $V(y)=0$ for $y_L < y < y_R$, and $V(y)=-V$ for
$y_R \le y \le w$ (see Fig.~\ref{pots}). At the points $y_L$ and
$y_R$ the potential jumps, giving rise to a singular electric field
only at $y_L$ and $y_R$ within the two-dimensional graphene sheet
pointing into the $y$ direction and being zero otherwise. $\pm V$
denotes the strength of the potential of the respective potential
steps. This special choice leads to a symmetric energy gap around the
Dirac points at $E=0$ if $y_L=w-y_R$.   

Due to the periodic boundary conditions in the $x$-direction, it is
convenient to take the Fourier-transforms  
$\xi(a,\vx) = e^{iqx} \xi(a,q,y)$, $\xi(b,\vx) = e^{iqx} \xi(b,q,y)$
and make the following ansatz for the wave function $\xi(a,q,y)$ 
in the three regions of $y$ 
\be
\xi(a,q,y)=\left\{
\begin{array}{l}
L_+ e^{ik_L y} + L_- e^{-ik_L y}, \hfill y\le y_L\\
M_+ e^{ik_M y} + M_- e^{-ik_M y}, \hfill y_L<y<y_R\\
R_+ e^{ik_R (y-w)} + R_- e^{-ik_R (y-w)}, \hfill y_R\le y 
\end{array}
\right.
\ee
where $k_{L,R}$ and $k_M$ are given by 
\be
k_{L,R} = \sqrt{(E \mp V)^2 - q^2}, \quad k_M = \sqrt{E^2 - q^2}. 
\ee
The roots are defined to be positive if $E$ is large
and the usual analytic continuation is taken otherwise.
$\xi(b,q,y)$ is then obtained from the Dirac equation as  
\be
\xi(b,q,y)=\left\{
\begin{array}{lll}
q^*_L L_+ e^{ik_L y} + q_L L_- e^{-ik_L y}, \hfill y\le y_L\\
q^*_M M_+ e^{ik_M y} + q_M M_- e^{-ik_M y}, \hfill y_L<y<y_R\\
q^*_R R_+ e^{ik_R (y-w)} + q_R R_- e^{-ik_R (y-w)}, \hfill y_R\le y
\end{array}
\right.
\ee
with
\be
 q_{L,R} = \frac{q+ik_{L,R}}{E\mp V}, \quad q_M = \frac{q+ik_M}{E}.
\ee
Here, $q_M^*$ is given by $q_M|_{k_M\to -k_M}$ even for imaginary
$k_M$ and we have $q_M^*  q_M=1$. The same applies to $q_{L,R}$. 

The six amplitudes $L_{\pm}$, $M_{\pm}$, and $R_{\pm}$ follow from the
normalization, from the boundary conditions 
\bea
 \xi(a,q,0) &=& L_+ + L_- =0  \nonumber \\
 \xi(b,q,w) &=& q_R^* R_+ + q_R R_- =0  \label{bc},
\eea
and from the matching conditions for $\xi(a,q,y)$ and $\xi(b,q,y)$
at $y_{L,R}$ 
\bea
 e_L L_+ + e_L^* L_-  &=& M_+' + M_-' \nonumber \\
 e_M M_+' + e_M^* M_-'  &=& e_R^* R_+ + e_R R_-  \nonumber \\
 q_L^* e_L L_+ + q_L e_L^* L_-  &=& q_M^* M_+' + q_M M_-' \nonumber \\
 q_M^* e_M M_+' + q_M e_M^* M_-' &=& q_R^* e_R^* R_+ + q_R e_R R_-  \label{mc}
\eea
with the abbreviations $M'_{\pm} = M_\pm e^{\pm i k_M y_L}$ and
\begin{equation}
 e_L = e^{ik_L y_L}\mathrm{,} \quad e_M = e^{ik_M(y_R-y_L)}\mathrm{,} 
\quad e_R = e^{ik_R(w-y_R)}. 
\end{equation}
The ``complex conjugates'' $e^*_L$, $e^*_M$, and $e^*_R$ are defined
as for the $q_L$, $q_M$, and $q_R$. Normalization of the wave function
demands a non-trivial solution of Eqs.~(\ref{bc}, \ref{mc}) and that
determines the energy. Furthermore, there are non-trivial solutions 
$k_L=0$ leading to $e_L=q_L=1, L_+ = - L_-, M_\pm =0, R_\pm=0$ and
similarly for $k_R=0$, $k_M=0$. These result in a zero wave
function and have to be excluded. 

After a straightforward calculation of the determinant of the $6 \times 6$
system, we get the condition for the eigenenergies in the form
$ f(E;q,V) = 0$ with a real $f$:
\bea
f(E;q, V) &=&
   \frac{1}{q_M-q_M^*} \frac{1}{q_R-q_R^*} \frac{1}{q_L-q_L^*}\times\nonumber \\
 && \Big\{q_M (1 - q_M^* q_R^*)(1 - q_M^* q_L^*) \;e_L e_M e_R -\nonumber \\
&& q_M^* (1 - q_M q_R)(1 - q_M q_L) \;e_L^* e_M^* e_R^* +  \nonumber \\
 && q_M (1 - q_M^* q_R)(1 - q_M^* q_L) \;e_L^* e_M e_R^* - \nonumber \\
&& q_M^* (1 - q_M q_R^*)(1 - q_M q_L^*) \;e_L e_M^* e_R -\nonumber \\
 && q_M (1 - q_M^* q_R)(1 - q_M^* q_L^*) \;e_L e_M e_R^* + \nonumber \\
&& q_M^* (1 - q_M q_R^*)(1 - q_M q_L) \;e_L^* e_M^* e_R -\nonumber \\
 && q_M (1 - q_M^* q_R^*)(1 - q_M^* q_L) \;e_L^* e_M e_R + \nonumber \\
&& q_M^* (1 - q_M q_R)(1 - q_M q_L^*) \;e_L e_M^* e_R^* \Big\}. 
\label{res}
\eea

This function is even in $V$ and for a symmetric arrangement,
$y_L=w-y_R$, we have $f(-E;q, V) = f(E;q, V)$. In the other valley,
the eigenvalues are determined by $f(E;-q,V) = 0$.
We denote the energy eigenvalues resulting from $f(E;q,V)$ by
$E_{s,n}(q)$, $s=\pm 1$, $n=0,1, \ldots, \infty$. 

In the absence of an external electric potential ($V=0$), those
wavefunctions corresponding to the eigenvalues close to $E=0$, i.e.,
$E_{s,0}(q\to\infty)$ with $s=\pm 1$, are localized along the 
edges.\cite{BF06a} With $M_{\pm}=L_{\pm}$ and $\kappa=\sqrt{q^2-E^2}$,
one gets 
\begin{eqnarray}
\xi(a,q,0<y<w)&=&-2M_+ \sinh(\kappa y)\\
\xi(b,q,0<y<w)&=&s2M_+ \sinh(\kappa (w-y)).
\end{eqnarray} 
Therefore, when the edge atoms at the left hand side are part of the
b-sublattice and the edge atoms at the right hand side part of the
a-sublattice, the wavefunction components of the b-sublattice are
concentrated on the left ($y=0$) for both energies $E_{s,0}(q)$,
while we find the opposite for the components of the a-sublattice,
which are concentrated at the right edge at $y=w$. The
edge states' width depends on the imaginary momentum $\kappa$.

\begin{figure}[t]
\begin{center}
\includegraphics[width=6.75cm]{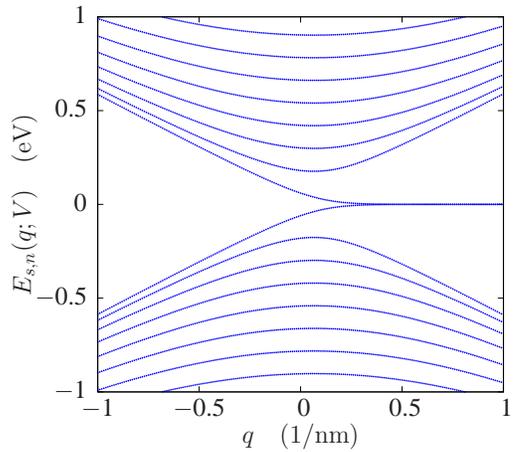}
\caption{(Color online) The energy spectrum $E_{s,n}(q;V)$ of a
  zig-zag graphene nanoribbon of width $w=15$\,nm in the absence of an
  external potential, $V=0$.}
\label{spectr00}
\end{center} 
\end{figure}

\begin{figure}[b]
\begin{center}
\includegraphics[width=6.75cm]{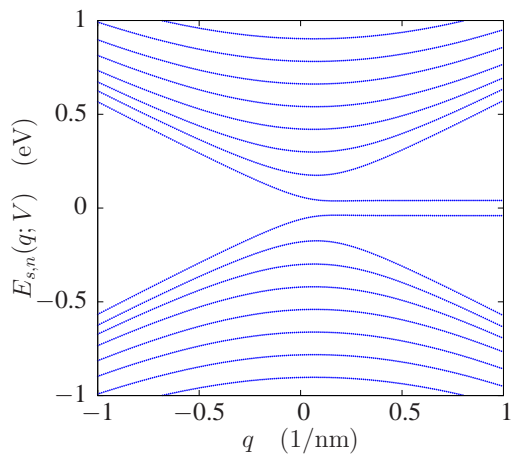}
\caption{(Color online) 
  The energy spectrum $E_{s,n}(q;V)$ of a zig-zag graphene
  nanoribbon of width $w=15$\,nm in the presence of external electric
  potentials $V=\pm 0.0403$\,eV. The width of the left and right
  strips ($y_L=w-y_R$) where the potentials are applied is $2w/5$.}
\label{spectr07}
\end{center} 
\end{figure}

\section{Results and Discussion\label{sectresult}}
\begin{figure}[t]
\begin{center}
\includegraphics[width=6.75cm]{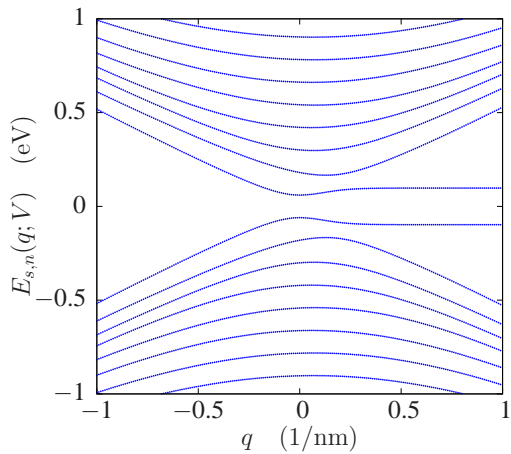}
\caption{(Color online) The energy spectrum $E_{s,n}(q;V)$ of a
  zig-zag graphene nanoribbon for external electric potentials $V=\pm
  0.0978$\,eV. The widths of ribbon and potential strips are as in
  Fig.~\ref{spectr07}.} 
\label{spectr17} 
\end{center}
\end{figure}

\begin{figure}[t]
\begin{center}
\includegraphics[width=6.75cm]{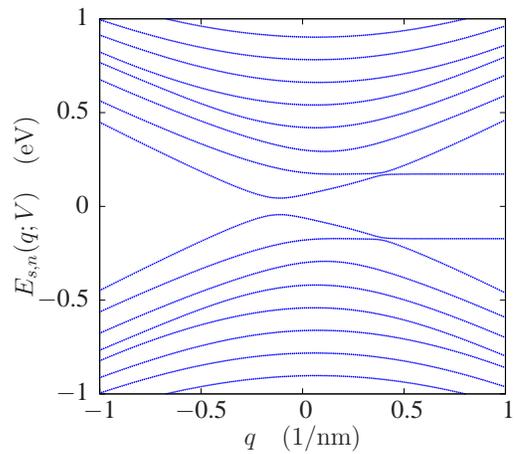}
\caption{(Color online) The energy spectrum $E_{s,n}(q;V)$ of a
  zig-zag graphene nanoribbon with electric potentials $V=\pm
  0.1725$\,eV. A closer inspection of the spectrum shows that the
  curves do not cross. The widths of ribbon and potential strips are
  as in Fig.~\ref{spectr07}.}  
\label{spectr30} 
\end{center}
\end{figure}

Now we turn to the discussion of the energy spectrum. 
In Figs.~\ref{spectr00}-\ref{spectr30}, we show a numerical evaluation
of $f(E;q,V)=0$. 
In graphene zig-zag nanoribbons, the bulk gap closes due to surface
states \cite{FWNK96,NFDD96} at the  
edges at $y=0$ and $y=w$. This is shown in Fig~\ref{spectr00}, 
where part of the energy spectrum $|E_{s,n}(q)|\le 1.0$\,eV as
obtained from (\ref{res}) is plotted for one valley and $V=0$.   
If we then apply a potential $\pm V$ that affects the eigenstates
located at the the zig-zag edges as described above, 
a gap opens proportional to $V$ until it reaches a maximum value 
determined by the width $w$ in $y$-direction. Here, we study the
symmetric case where $y_L=w-y_R$. 
For a finite electric potential $\pm V$, the magnitude of the spectral
gap depends on the width of the electrodes, $y_L = w-y_R$,  as
long as $y_L$ is smaller than the effective width $w_e\approx q^{-1}$
of the edge state. Therefore, a much stronger $V$ must be applied to
achieve the same spectral gap if $y_L < w_e$. The energy gap increases
with $y_L$ and saturates at $y_L \gtrsim w_e$ for which the energy
spectrum becomes independent of $y_L$. The latter situation is realized
in Figs.~\ref{spectr07}, \ref{spectr17}, and  \ref{spectr30}, where
the steps that define the width of the potential strips are chosen to
be at 
$y_L=0.4\,w$ and $y_R=0.6\,w$, respectively. If we continue to increase
$V$, the gap closes again (see Fig.~\ref{spectr30}) although the edge
states, which can always be identified by their nearly dispersionless
eigenenergies, move further apart because they are strongly affected
by the external potential. With increasing $V$, the transition point
where the bulk state transform into edge states moves to larger $q$.     

The energy spectrum at $q=0$ can be found for arbitrary potential
widths $y_L$ and $w-y_R$ by putting in (\ref{res}) $q_L=q_M=q_R=i$.
We get
\begin{equation}
E_{s,n}(q=0)=s\frac{\pi}{w}(n+\frac{1}{2})-V\frac{w-y_L-y_R}{w}.
\label{qnull}
\end{equation}
For $n \gtrsim 1$, the minima of the electron subbands and the maxima
of the hole subbands appear not at $q=0$ but at $q\simeq 1/w$. Also,
for $2V<\pi\hbar v_{\rm F}/w$ and increasing $q$, the electronic
states at $E_{s,0}(q)$ start to become almost dispersionless surface
states close to the momentum $q\simeq 1/w$. The corresponding
eigenenergies can be calculated in the range $1/w < q < 1/a_0$, where
$a_0$ is the lattice constant of the underlying lattice model. 
A careful evaluation of (\ref{res}) in the limit $q\to\infty$ yields
for the upper ($s=1$) and lower ($s=-1$) state    

\begin{equation}
E_{s,0}(1/w<q<1/a_0)\propto \left\{
\begin{array}{lll}
s|V|(1-e^{-2qy_L}) &\?\mathrm{;}\?& sV>0\\
s|V| (1-e^{-2q(w-y_R)}) &\?\mathrm{;}\?& sV<0.
\end{array}\right. 
\end{equation}

The appertaining density of states can also be estimated in this
limit. It drops from large values at $\pm V$ down to zero at the
Dirac point. For $V=0$, however, we get $E_{s,0}(q)=s2q e^{-qw}$ 
and so the corresponding density of states behaves in the limit of
$|Ew| \to 0$ as $\sim |Ew|^{-1}$. 

Please note that in the Dirac approximation, the almost dispersionless
states at $E_{\pm 1,0}(q)$ seem to be true solutions for all
$q\to\infty$. However, if one notices that this model originates from
a lattice model, one sees that the nearly dispersionless states merely
connect the two valleys at $\vk=+\vK$ and $\vk=-\vK$. 
That means, the Dirac model needs to be supplied with a cut-off
($q_{\rm max} \approx 1/a_0$) in order to properly describe the
asymptotic ($E\sim 0$) region of the tight-binding (TB) model.

\section{Transport, influence of disorder and conclusions\label{sectdis}}
\begin{figure}
\begin{center}
\includegraphics[width=7.0cm]{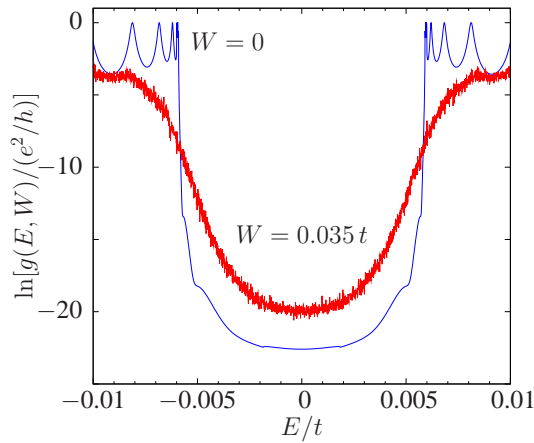}
\caption{(Color online) The logarithm of the two-terminal conductance
  $g(E)$ vs.\ energy $E$ of a perfect (blue curve) narrow 
  zig-zag graphene lattice of 15\,nm width in the presence an
  effective electric potential $V/t=\pm 0.006$ applied  along the
  edges showing a transport gap $\Delta E=2V$. The red curve reflects
  the influence of additional edge disorder $W/t=0.035$ (see text below).}     
\label{2termg}
\end{center}
\end{figure}

To support our finding of an induced spectral gap obtained within 
the continuum Dirac-model and to see its influence on the transport
properties, we calculate numerically the two-terminal conductance 
$g(E)$ of narrow graphene zig-zag ribbons applying a transfer-matrix 
method within a TB-lattice model.\cite{SM08a} Here, semi-infinite
leads are attached to both ends of the finite nanoribbon and the
electric conductance is defined as usual via the transmission through
the entire system (see Ref.~{\onlinecite{SM08a}} and references
therein for details). As an example, the logarithm of the energy
dependent $g(E)$ is shown in Fig.~\ref{2termg} for a clean 15\,nm
narrow nanoribbon with an electric potential $V/t=\pm 0.006$ applied
along the edges leading to a transport gap around the Dirac
point. Here, $t\approx 2.7$\,eV is the nearest-neighbor hopping term
in graphene. The conductance exhibits sharp resonances with maxima
$g(E)\simeq e^2/h$ and decreases down to very small values $\sim
10^{-10}e^2/h$ for energies between $\pm 0.006\,t$. The latter is due
to tunneling and depends on the length of the sample. 

\begin{figure}[t]
\begin{center}
\includegraphics[width=6.5cm]{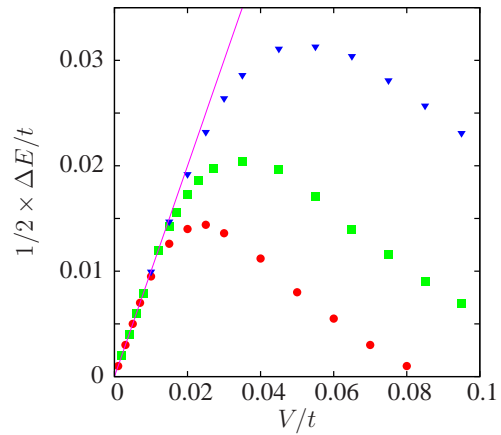}
\caption[]{(Color online) The transport gap $\Delta E$ vs.\ external
  potential $V$ obtained numerically from the energy dependent
  two-terminal conductance $g(E)$ of narrow zig-zag graphene lattices
  having widths 10\,nm ($\blacktriangledown$), 15\,nm (\blx), and
  22\,nm ($\bullet$), respectively. The straight line is $\Delta E=2V$
  and $t\approx 2.7$\,eV is the tight-binding hopping energy.}   
\label{tgap}
\end{center}
\end{figure}

Fig.~\ref{tgap} shows the dependence of the induced transport gap
$\Delta E$ on the effective external potential $V$, and that nicely
confirms the linear behavior of our analytical results for small $V$.  
The transport gap data plotted vs.\ potential in Fig~(\ref{tgap}) can
be re-scaled by the respective ribbon widths. The outcome of this
procedure is, within the given uncertainty of the data, a single
fitting curve for all data (not shown). 
This result can be understood from an evaluation of (\ref{res}) where
one finds for $y_L=w-y_R$ that the maximal gap appears always at the
Dirac point $q=0$. Therefore, Eq.~(\ref{qnull}) can be used to see
that $2E_{1,0}w=\pi$ which together with the observed $1/2 \Delta E=V$
relation gives the scaling behavior mentioned above.   
The maximal transport gap observed in the lattice model agrees within
the numerical uncertainty with the spectral gap of both the finite
lattice model and the continuum model where $L\to\infty$ was assumed.  
Recovering the units, we get from (\ref{qnull}) and $\hbar v_F=3/2 t
c$, where $c=1.42\times 10^{-10}$\,m is the carbon-carbon bond length,
that the maximal spectral gap (for $y_L=w-y_R$) follows the relation
\begin{equation}
E_{\rm max}(w)=\frac{\pi}{w}\hbar v_{\rm F}=\frac{3}{2} \frac{\pi}{w}tc,
\end{equation}  
leading to $E_{\rm max}\approx 0.12$\,eV for $W=15$\,nm.
We also find in our numerical work that replacing
the assumed piece-wise constant effective electric potentials by more
realistic smooth potential steps does not modify the results.  

Finally, to check the robustness of this proposed gap opening
mechanism against edge disorder, which may arise, e.g., through edge
passivation by randomly placed hydrogen atoms that is known to
stabilize the edges of pristine zig-zag nanoribbons 
considerably,\cite{WSSLM08} we apply a random disorder 
potential along the border of the ribbon. The eigenvalues are
obtained by standard diagonalization of the Hamilton matrix for
$L_x\times L_y$ graphene zig-zag ribbons described by the
TB-Hamiltonian defined on a bricklayer lattice\cite{SM08a} with sites
$\vvr$ and nearest neighbor distance $a$
\begin{equation}  
{\cal H}=\sum_{\vvr} \epsilon_{\vvr} c_{\vvr}^{\dagger}c_{\vvr}^{}-
t\sum_{\langle \vvr\ne\vvr'\rangle} c_{\vvr}^{\dagger}c_{\vvr'}^{},
\end{equation}
where $\langle \vvr\ne\vvr'\rangle$ are pairs of those neighboring
sites that are mutual connected on the bricklayer. 
The  disorder potentials $\epsilon_r$ are uncorrelated random numbers
that are non-zero only at the outer sites (different sublattice on
left and right edge) along the zig-zag edges of the nanoribbon and
uniformly distributed between $\pm W$, where $W$ denotes the disorder
strength.   

\begin{figure}
\begin{center}
\includegraphics[width=7.5cm]{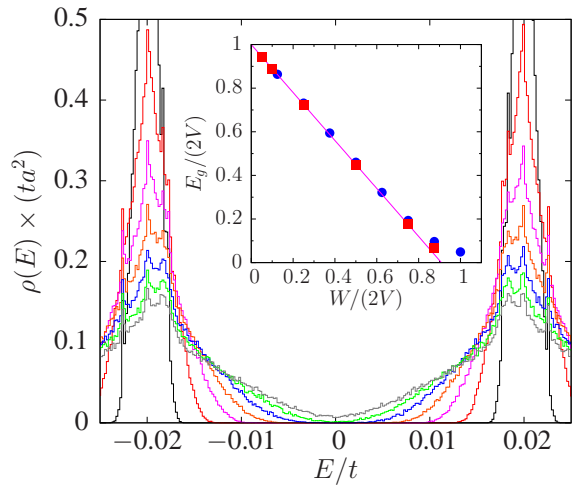}
\caption[]{(Color online) The density of states $\rho(E)$ of graphene
  zig-zag nanoribbons ($L_x/a=720$, $L_y/a=72$) with external edge
  potentials $V/t=\pm 0.02$ in the energy interval $-0.025\le E/t \le
  0.025$ around the Dirac point. One identifies the disorder
  broadening of the peaks at $E/t=\pm 0.02$, which belong to the
  eigenstates located 
  along the ribbon edges, for disorder strengths $W/t=$ 0.005
  (black narrow peaks), 0.01 (red), 0.015 (magenta), 0.02 (orange),
  0.025 (blue), 0.03 (green), and 0.035 (grey broadest peaks). The
  mean spectral gap $E_g$ decreases linearly with increasing disorder
  strength $W$. This is shown in the inset ($\bullet$) together with
  additional data for a system of size $L_x/a=960$, $L_y/a=48$ and
  $V/t=0.01$ (\blx) which all scale onto a single curve
  $E_g=2V-1.1\,W$ almost until the gap closes.}        
\label{disorder}
\end{center}
\end{figure}

The resulting density of states (DOS) of a 15\,nm zig-zag nanoribbon with  
$L_x/a=720$, $L_y/a=72$, and $V/t=0.02$, averaged over 1000 disorder
realizations, shows a broadening of the DOS-peaks around $E/t=\pm
0.02$ that originate from the edge states (see Fig.~\ref{disorder}).  
With increasing disorder potential strength $W/t=$ 0.005, 0.01, 0.015,
0.02, 0.025, 0.03, and 0.035, the spectral gap decreases linearly. 
This is seen in the inset of Fig.~\ref{disorder}, where the
above results are shown together with additional data from a system of
size $L_x/a=960$, $L_y/a=48$ (10\,nm ribbon), and $V/t=0.01$. 
All data points collapse onto the function $E_g=2V-1.1 W$, which was
also observed for other ribbon sizes having widths larger than 7\,nm. 
Therefore, a gap should remain open in experiments when $V$ is tuned
to be larger than $0.55\times W$, where $W$ is usually fixed by the
sample dependent intrinsic edge disorder. Here, the mean
energy gap $E_g=1/N_r \sum_{i}^{N_r} (E_{i}^{+}-E_{i}^{-})$ is defined
as the  ensemble averaged difference between the smallest positive and
the largest negative eigenvalue averaged over $N_r$ realizations.      
For general disorder with box-probability density-distributions
$P(\epsilon)$, 
we find a gap closing relation $E_g=2V-\kappa \Gamma_2^{1/2}$, where
$\Gamma_2=\int_{-\infty}^{\infty}\epsilon^2 P(\epsilon) d\epsilon$ is
the second moment of the disorder distribution and $\kappa=1.9$ is an
empirical constant.   

The influence of edge disorder on the logarithm of the two-terminal
conductance, $\ln g(E,W)$ averaged over 100 realizations, is
shown in Fig.~\ref{2termg} for a finite lattice of width $L_y/a=72$
and length $L_x/a=720$. Due to the random edge potentials, the sharp
conductance resonances of the clean sample are smoothened out. Yet, a
transport gap remains visible even in the case of strong disorder
$W=0.035\,t$ when the spectral gap has completely vanished but $g(E)$
still drops six orders of magnitudes from about $g\simeq 0.1\,e^2/h$ at
$E=0.008\,t$ to $g=10^{-7}e^2/h$ at $E=0$. This means that the almost
one-dimensional edge states of the clean sample become Anderson
localized in the presence of sufficient edge disorder. This
notion has been corroborated by an investigation of the respective
eigenstates and by calculations of the length and disorder dependence
of $g(E)$. Previous studies have reached similar conclusions for 
nanoribbons with rough edges.\cite{EZHH08,MCL09}  
    
In conclusion, we have shown that the application of external electric
potentials, covering the area of the electronic edge states that are
located along the zig-zag edges of a graphene nanoribbon, can open a
tunable spectral gap. Thus, one can convert the metallic behavior into
a semiconducting one. For small potentials, the gap increases linearly
with the potential strength, reaches a ribbon-width-dependent maximum
$\pi\hbar v_{\rm F}/w$ ($\approx$ 0.12\,eV for $w=$15\,nm) and closes
again with further increasing electric potentials. The origin of this
effect comes from the sensitivity of the spinorial edge states to
electric potentials. Applying distinct external biases to the
left and right edge state leads to a different shift of the almost
dispersionless edge energies as long as they are not pinned to the
Fermi level. Using electric potentials of opposite sign causes the
largest energy gap possible. The disorder effects, which may be
due to atoms and molecules that saturate the dangling-bonds along the
zig-zag edges in real samples, are found to reduce the spectral gap.
The latter remains, however, finite as long as $W<1.82 V$, where $W$
is a measure of the disorder strength and $V$ the applied effective
electric potential. For even larger disorder strengths, a transport
gap is still present allowing for reasonable on-off-rations 
for the electric current.  
Future experiments will show whether the present results of single
particle physics are sufficient for the description of a gap 
opening by external potentials in graphene zig-zag nanoribbons or
if theories that emphasize edge magnetism
\cite{SCL06,SCL06a,JPM09} due to $e$-$e$ interactions have to be
applied.    

\textit{Note added:} During the review procedure, we became aware of a
recent paper \cite{BS10} by Bhowmick and Shenoy that addresses a
spectral gap opening induced by external $\delta$-like potentials
placed along the edges of graphene zig-zag ribbons. This specific
potential choice represents a special case contained in our model.


\end{document}